\documentclass[showpacs,superscriptaddress,twocolumn,prb]{revtex4}%
\usepackage{amsfonts}
\usepackage{amsmath}
\usepackage{amssymb}
\usepackage{graphicx}%
\begin{document}
\title{Spin current in the Kondo lattice model}
\author{Shun-Qing Shen}
\affiliation{Department of Physics, The University of Hong Kong, Pokfulam Road, Hong Kong, China}
\author{X. C. Xie}
\affiliation{Department of Physics, Oklahoma State University, Stillwater, OK 74078 and
International Center for Quantum Structure, Chinese Acedemy of Sciences,
Beijing, China}
\date{January 4, 2004}

\begin{abstract}
By using the projection operator technique it is observed that the strong
Hund's rule coupling and s-d interaction in transition metal elements may lead
to an effective coupling between the spin current and spin chirality. As a
result, the spin chirality can be regarded as a driving force to produce a
spin current. The spin current may give rise to a novel type of field acting
on the spins. A spin battery is designed based on the interactions between the
spin current and spin chirality.

\end{abstract}
\maketitle

Spin-dependent effects, such as giant or colossal magnetoresistance and
magnetization switching, arise from the interaction between spin of charge
carrier and an applied external magnetic field. In spintronics, instead of
charge, electron spin carries information and it is possible that capability
and performance can be enhanced in spintronics
devices.\cite{Wolf01,Ziese01,Awschalom02} Possible applications and
fundamental science involved make the study of spin-dependent transport an
intensive field in condensed matter physics. Recently, several ways to produce
a pure spin current are proposed, such as anomalous Hall effect in
ferromagnetic metals\cite{Hirsch99,Zhang00}, ferromagnetic resonance
\cite{Berger99,Brataas02}, and the spin current in a spin spiral
state.\cite{Shen97, Konig01} In this paper, we study spin current in the Kondo
lattice model, a simplified model containing essential physics for many
ferromagnetic metals or semiconductors. We find a new type of interaction
between spin current and spin chirality (defined later) in the strong Hund
coupling limit. This indicates that the spin chirality can be regarded as a
driving force to produce spin current, similar as an electric field does to
the electric current. Furthermore, the spin current induces a new type of an
effective magnetic field. Based on the interaction between spin current and
spin chirality, we propose a scheme to design a spin battery. As a
demonstration, we study the spin transport in the well known spiral state to
bring out the above mentioned spin-current effects.

The Hund's rule coupling and s-d exchange interaction are crucial in
transition metal elements. It plays a key role in the formation of metallic
ferromagnetism in colossal magnetoresistance materials
\cite{Zener51,Anderson55,Kubo72} and diluted magnetic
semiconductors.\cite{Ohno98,Dietl00} In this paper we focus on doped
transition metal oxides and/or ferromagnetic semiconductors. The minimal model
for these materials is the Kondo lattice model,%
\begin{equation}
H=-t\sum_{\left\langle n,m\right\rangle ,\sigma}c_{n,\sigma}^{\dagger
}c_{m,\sigma}-\frac{J}{2}\sum_{n,\sigma\sigma^{\prime}}\mathbf{S}_{n}\cdot
\tau_{\sigma\sigma^{\prime}}c_{n,\sigma}^{\dagger}c_{n,\sigma^{\prime}},
\end{equation}
where $c_{n,\sigma}^{\dagger}$ and $c_{n,\sigma}$ are the creation and
annihilation operators for conduction electrons, $\mathbf{S}_{n}$ is the spin
operators for the localized spin, and $\tau$ are the Pauli matrices and
$\tau_{0}$ is the $2\times2$ identity matrix. The summation over $\left\langle
n,m\right\rangle $ runs for the nearest neighbor pairs of lattice sites. In
this paper we will set $\hbar=c=1$ and the lattice space is $a$ and is set to
unit.\cite{Scalapino93} For the Hund's rule coupling, $J$ is always positive.
The model has been studied extensively.\cite{Tsunetsugu97} The spin current
operator along $\alpha$ axis in the tight binding approximation is defined as
$t\hbar\mathbf{I}_{nn+\alpha}^{s}$ and
\begin{equation}
\mathbf{I}_{nn+\alpha}^{s}=-\frac{i}{2}\sum_{\sigma,\sigma^{\prime}}\left(
c_{n,\sigma}^{\dagger}\tau_{\sigma\sigma^{\prime}}c_{n+\alpha,\sigma^{\prime}%
}-c_{n+\alpha,\sigma}^{\dagger}\tau_{\sigma\sigma^{\prime}}c_{n,\sigma
^{\prime}}\right)  ,
\end{equation}
where $\alpha(=x,y,z)$ are vectors pointing to the nearest neighbor sites,
which have one longitudinal and two transverse components. The spin polarized
currents and electric current become equal in a fully polarized ferromagnetic system.

\begin{figure}
[ptb]
\begin{center}
\includegraphics[
height=1.9138in,
width=2.3125in
]%
{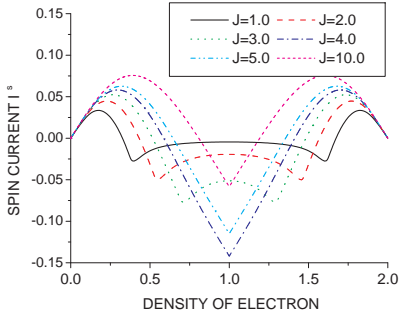}%
\caption{Spin current via density of conduction electron in the spin spiral
state of $\theta=\pi/2$ and $\varphi=0.2\pi.$}%
\label{current}%
\end{center}
\end{figure}

\begin{figure}
[ptb]
\begin{center}
\includegraphics[
height=1.9761in,
width=2.3099in
]%
{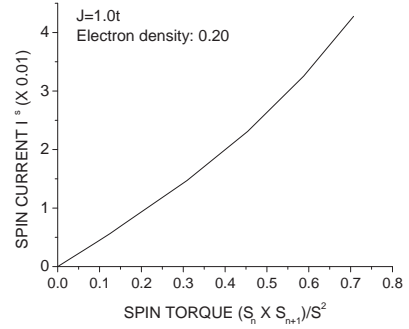}%
\caption{Spin current via spin chirality at $J=1.0$ with the density of
conduction electron 0.20.}%
\label{torque}%
\end{center}
\end{figure}

Before we discuss the spin transport of Eq.(1) in an arbitrary dimension, let
us first to illustrate the physics by considering the well known spin spiral
state in one-dimensional system. In general, the spin spiral state may not be
the ground state of Eq.(1), however, it can be stabilized by external
environments, such as by applying external fields along different directions
at the ends of the system.\cite{Spiral} Our purpose here is to demonstrate the
consequence of the spin spiral state. We parameterize $\mathbf{S}_{n}$ by the
polar angles $\theta_{n}$ and $\varphi_{n},$ i.e., making the classical spin
approximation. The spin spiral state is defined by $\mathbf{S}_{n}/S=\left(
\sin\theta\cos n\varphi,\sin\theta\sin n\varphi,\cos\theta\right)  .$ The spin
chirality defined as the cross times between the two neighboring spins:
$\left\langle \mathbf{S}_{n}\times\mathbf{S}_{n+1}\right\rangle _{z}=S^{2}%
\sin^{2}\theta\sin\varphi.$ In the k- space, we introduce a spinor
operator\ $\Phi^{\dagger}(k)=\left(  c_{ka+\varphi/2,\uparrow}^{\dagger
},c_{ka-\varphi/2,\downarrow}^{\dagger}\right)  .$ The mean field Hamiltonian
is written as%
\begin{equation}
H=\sum_{k}\Phi^{\dagger}(k)H(k)\Phi(k),
\end{equation}
where $H(k)=-g_{0}\tau_{0}-g_{x}\tau_{x}-g_{z}\tau_{z}$ with $g_{0}=\mu
+2t\cos\frac{\varphi}{2}\cos ka;$ $g_{x}=\frac{J}{2}\sin\theta;$ $g_{z}%
=\frac{J}{2}\cos\theta-2t\sin\frac{\varphi}{2}\sin ka.$ $\mu$ is the chemical
potential. The single particle Green function in a $2\times2$ matrix form is
given by
\begin{align}
G(k,i\omega_{n})  &  =(i\omega_{n}\tau_{0}-H(k))^{-1}\nonumber\\
&  =\frac{\left[  (i\omega_{n}+g_{0})\tau_{0}-g_{x}\tau_{x}-g_{z}\tau
_{z}\right]  }{\left(  i\omega_{n}-\omega_{+}+\mu\right)  \left(  i\omega
_{n}-\omega_{-}+\mu\right)  },
\end{align}
where $\omega_{n}=(2n+1)\pi kT$ with an integer $n$ and $T$ is the
temperature. The two branches of spectra are%
\begin{equation}%
\begin{array}
[c]{l}%
\omega_{\pm}(k)=-2t\cos\frac{\varphi}{2}\cos ka\\
\pm\left[  \left(  \frac{J}{2}\cos\theta-2t\sin\frac{\varphi}{2}\sin
ka\right)  ^{2}+\frac{J^{2}}{4}\sin^{2}\theta\right]  ^{1/2}.
\end{array}
\end{equation}
In the k-space, the expectation value of the longitudinal spin current
operator is written\ in terms of the Green's function,%
\begin{equation}
\left\langle I_{z}^{s}\right\rangle =\frac{1}{N}\sum_{k}\left[  A_{+}%
n_{F}(\omega_{+}-\mu)-A_{-}n_{F}(\omega_{-}-\mu)\right]  , \label{iz}%
\end{equation}
where
\begin{align*}
A_{\pm}(k)  &  =\pm\sin\frac{\varphi}{2}\cos ka\\
&  +\frac{t\sin\varphi\sin^{2}ka-\frac{J}{2}\cos\theta\cos\frac{\varphi}%
{2}\sin ka}{\sqrt{\left(  \frac{J}{2}\cos\theta-2t\sin\frac{\varphi}{2}\sin
ka\right)  ^{2}+\frac{J^{2}}{4}\sin^{2}\theta}}.
\end{align*}
We also calculated the electric current $I_{z}^{c}$ and found that $I_{z}%
^{c}=0$. Thus in the spiral state there exists a spin current without a charge
current. In Fig. \ref{current} we plot spin current as a function of electron
density in the state with $\theta=\pi/2,$ and $\varphi=0.2\pi$ for different
$J$. We see that a larger $J$ produces a higher spin current. For a finite
$J$\ the current may change direction with change of density. In Fig.
\ref{torque} we plot the relation between spin current and spin chirality at
the density of $0.2$. The spin current increases with the spin chirality.%

%

In the large $J$ limit, the problem can be simplified. An equivalent
Hamiltonian of Eq.(1) or so-called double exchange model in the large $J$
limit is given by\cite{Hartman96}%
\begin{equation}
H_{de}=-\sum_{\left\langle n,m\right\rangle }t_{nm}\alpha_{n}^{+}\alpha_{m},
\label{de}%
\end{equation}
with $t_{nm}=$ t$\left[  \cos\frac{\theta_{n}}{2}\cos\frac{\theta_{m}}{2}%
+\sin\frac{\theta_{n}}{2}\sin\frac{\theta_{m}}{2}e^{-i(\varphi_{n}-\varphi
_{m})}\right]  $, and $\alpha_{n}=\cos(\theta_{n}/2)(1-n_{n\downarrow
})c_{n\uparrow}+\exp[i\varphi_{n}]\sin(\theta_{n}/2)(1-n_{n\uparrow
})c_{n\downarrow}$ and satisfies the anticommutation relation, $\left[
\alpha_{k},\alpha_{k^{\prime}}^{\dagger}\right]  =\delta_{k,k^{\prime}}.$ The
energy excitations and their eigenstates can be obtained from the diagonalized
Hamiltonian,
\begin{equation}
H_{de}=-2t\sum\left[  \cos^{2}\frac{\theta}{2}\cos ka+\sin^{2}\frac{\theta}%
{2}\cos(ka+\varphi)\right]  \alpha_{k}^{\dagger}\alpha_{k},
\end{equation}
where $\alpha_{k}$ is the Fourier transform of $\alpha_{n}.$ In the case when
the spins of conduction electrons are frozen by the localized spins, and the
quasiparticles become spinless fermions. The two Fermi momenta are $k_{\pm
}^{F}a=-k_{0}\pm\rho\pi$ where $k_{0}=\arctan[\sin^{2}\theta\sin\varphi
/(\cos^{2}\theta+\sin^{2}\theta\cos\varphi)]$ and $\rho$ is the ratio of the
number of conduction electrons to the number of total lattice sites and
$\rho/a$ is the density of conduction electrons$.$ $k_{0}$ is nonzero when the
order parameter for the spin spiralling is nonzero. Its lowest energy state is
$\left\vert GS\right\rangle =\prod\limits_{k_{F}^{-}\leq k\leq k_{F}^{+}%
}\alpha_{k}^{\dagger}\left\vert 0\right\rangle .$ The spin current in the
state can be calculated from $\mathbf{I}_{z}^{s}=\sum_{n}\left\langle
GS\right\vert \mathbf{I}_{nn+1}^{s}\left\vert GS\right\rangle /N.$ We have an
analytical expression for spin current,%
\begin{equation}
I_{z}^{s}=t\hbar\frac{\sin\rho\pi}{\pi}\frac{\sin^{2}\theta\sin\varphi}%
{\sqrt{1-\sin^{2}\theta\sin^{2}(\varphi/2)}}. \label{spin-current}%
\end{equation}
The result is equal to what we obtain by taking the large limit in
Eq.(\ref{iz}). In the point of view of the Berry phase the spin spiral state
acquires a constant Berry phase in the spin-dependent renormalized factor in
the Hamiltonian (Eq. \ref{de}), i.e., $t_{nn+1}=\left\vert t_{nn+1}\right\vert
e^{-i\delta\varphi}.$ The non-zero phase $\delta\varphi$ drives a spin current.

By analogy with electric field and resistance, we introduce the concept of
\textit{spin current resistance} or its reciprocal, \textit{spin conductance}.
The role of $\mathbf{S}_{n}\times\mathbf{S}_{m}$ here is analogous to the
local electric field. In the spin spiral state the spin chirality is non-zero,
$\left\langle \mathbf{S}_{n}\times\mathbf{S}_{n+1}\right\rangle _{z}=S^{2}%
\sin^{2}\theta\sin\varphi.$ In Eq.(\ref{spin-current}) the current is
proportional to the spin chirality $\left\langle \mathbf{S}_{n}\times
\mathbf{S}_{n+1}\right\rangle _{z},$ and can be written in a compact form
$I_{z}^{s}=c\cdot\left\langle \mathbf{S}_{n}\times\mathbf{S}_{n+1}%
\right\rangle _{z}/S^{2}$ by introducing a spin conductance
\begin{equation}
c=\sin\rho\pi/\left[  \pi\sqrt{1-\sin^{2}\theta\sin^{2}(\varphi/2)}\right]
\label{conductance}%
\end{equation}
in a large J limit. The spin conductance is determined by the filling of
electrons and the magnetic structure. Just like an electric conductance, the
spin conductance reflects intrinsic properties of spin transport in materials.
It depends on the spin-dependent scattering mechanism of electrons. In Eq.
(\ref{conductance}), for the case of $\rho\rightarrow0$ or $\rightarrow1,$ c
is proportional to the density of electrons $\rho$ or the density of holes
$1-\rho$. For a small spin chirality $\sin^{2}\theta\sin^{2}(\varphi
/2)\rightarrow0,$ the spin conductance aprroaches to $[\sin\rho\pi]/\pi$,
which is proportional to the averaging kinetical energy of the charge carriers
in this tight binding theory. From the numerical results in Figs. 1 and 2, we
deduce that the spin conductance also depends on $J$. It becomes more
complicated for the finite $J$ cases. Similarly, we can also introduce a spin
voltage, which is also proportional to the length along the spin spiraling
path, $V_{spin}=\int\left\langle \mathbf{S}_{n}\times\mathbf{S}_{n+1}%
\right\rangle \cdot d\mathbf{l/}S^{2}$ where $d\mathbf{l}=\mathbf{r}%
_{n+1}-\mathbf{r}_{n}$ is the difference of two position vectors of
neighboring spins. $V_{spin}$ is an important parameter characterizing the
ability to produce a spin current.

Thus far we have introduced several important concepts, such as spin chirality
or spin current resistance, through the example of the classical spiral state.
To determine the microscopic mechanism to produce a spin current, we go back
to the Kondo lattice model (Eq.(1)) in the large Hund's rule coupling limit
and address the quantum spin case. The following discussions are not limited
to one-dimensional systems and the conclusion is independent of
dimensionality. According to the Hund's rule coupling, the energy of empty or
double occupancy of conduction electrons is zero. For a single occupancy the
conduction electron and localized spin can form a spin $S+1/2$ state with
$-JS/2$ and a spin $S-1/2$ state with $+J(S+1)/2$. In the large $J$ limit the
spin $S-1/2$ state and double occupancy should be excluded as they have much
higher energy than the spin $S+1/2$ state. This process can be realized with
the help of the projection technique.\cite{Fulde97} For that, we introduce the
spinor $\phi_{n}^{+}=\left(  (1-n_{n,\downarrow})c_{n,\uparrow}^{\dagger
},(1-n_{n,\uparrow})c_{n,\downarrow}^{\dagger}\right)  .$ The dressed
operators $(1-n_{n,-\sigma})c_{n,\sigma}^{\dagger}$, instead of $c_{n,\sigma
}^{\dagger}$, are used here to avoid the double occupancy due to the strong
coupling. The projection operator for the spin $S+1/2$ state is defined as
\begin{equation}
\mathcal{P}_{n}^{+}=\phi_{n}^{+}\mathbf{P}_{n}^{+}\phi_{n},
\end{equation}
where $\mathbf{P}_{n}^{+}=[\mathbf{S}_{n}\cdot\tau+(S+1)\tau_{0}]/(2S+1)$. The
operator $\mathcal{P}_{n}^{+}$ forces the electron spin at site $n$ to be
fully polarized with the localized spin at the same site. The projection
operator for non-occupancy is $\mathcal{P}_{n}^{e}=(1-n_{n,\downarrow
})(1-n_{n,\uparrow}).$ The total projection operator to exclude the spin S-1/2
single occupancy and double occupancy is $\mathcal{P}=\prod\limits_{n}\left(
\mathcal{P}_{n}^{e}+\mathcal{P}_{n}^{+}\right)  .$ Therefore the
\textit{equivalent} Hamiltonian in the strong Hund's rule coupling limit is
expressed as\cite{Kubo72,Shen98}%
\begin{equation}
H_{eff}=\mathcal{P}H\mathcal{P}=-t\sum_{nm}\left[  \phi_{n}^{+}\mathbf{P}%
_{n}^{+}\mathbf{P}_{m}^{+}\phi_{m}+h.c.\right]  , \label{eff}%
\end{equation}
where a constant is omitted in the last step. In the large S approximation the
Hamiltonian of (Eq.\ref{eff}) becomes the double exchange model in
Eq.(\ref{de}). Using a mathematical identity%
\begin{equation}
\left(  \mathbf{S}_{n}\cdot\tau\right)  \left(  \mathbf{S}_{m}\cdot
\tau\right)  =\mathbf{S}_{n}\cdot\mathbf{S}_{m}\tau_{0}+i\left(
\mathbf{S}_{n}\times\mathbf{S}_{m}\right)  \cdot\tau,
\end{equation}
the Hamiltonian can be rewritten as\cite{Note}%
\begin{align}
H  &  =-\frac{\left(  S+1\right)  ^{2}t}{\left(  2S+1\right)  ^{2}}\sum
_{nm}k_{nm}\nonumber\\
&  -\frac{t}{\left(  2S+1\right)  ^{2}}\sum_{nm}k_{nm}\mathbf{S}_{n}%
\cdot\mathbf{S}_{m}\nonumber\\
&  -\frac{\left(  S+1\right)  t}{\left(  2S+1\right)  ^{2}}\sum_{nm}\left(
\mathbf{S}_{n}+\mathbf{S}_{m}\right)  \cdot\mathbf{M}_{nm}\nonumber\\
&  +\frac{t}{(2S+1)^{2}}\sum_{nm}\left(  \mathbf{S}_{n}\times\mathbf{S}%
_{m}\right)  \cdot\mathbf{I}_{nm}^{s}, \label{effective}%
\end{align}
where $k_{nm}=\phi_{n}^{+}\tau_{0}\phi_{m}+\phi_{m}^{+}\tau_{0}\phi_{n}$ and
$\mathbf{M}_{nm}=\phi_{n}^{+}\tau\phi_{m}+\phi_{m}^{+}\tau\phi_{n}.$ The first
term is the kinetic part for conduction electrons excluding double occupancy,
like the $t$-term in the $t-J$ model. The second term contains an effective
Heisenberg-type spin-spin exchange interaction. Approximately, the effective
coupling is $J_{ferro}=-t\left\langle k_{nm}\right\rangle /(2S+1)^{2}$,
proportional to the kinetic energy in the sense of the mean field
approximation. The kinetic energy $-t\left\langle k_{nm}\right\rangle $ in the
tight binding approximation is always negative if the electron filling
$\rho<1$ such that the effective coupling is ferromagnetic. This is consistent
with the double exchange picture.\cite{Furukawa95, Millis95} The third term is
the interaction between local spin $\mathbf{S}_{n}$ and the vector
$\mathbf{M}_{nm}$. For a paramagnetic state, $\left\langle \mathbf{M}%
_{nm}\right\rangle $ is equal to zero, but in a fully polarized state, its
value is equal to $\left\langle k_{nm}\right\rangle ,$ i.e. proportional to
the kinetic energy. Among the four terms the most important observation is
from the last term: \textit{it contains an interaction between the spin
current operator }$\mathbf{I}_{nm}^{s}$\textit{ and the spin chirality
}$\mathbf{S}_{n}\times\mathbf{S}_{m}$\textit{. }By analogy with the coupling
between the electric current and electric field it is straightforward to
understand why a non-zero spin chirality induces a pure spin current. Although
we cannot derive an analytic expression for a finite $J$, it is expected that
the mechanism to produce a spin current remains the same.

\begin{figure}
[ptb]
\begin{center}
\includegraphics[
height=1.433in,
width=2.2494in
]%
{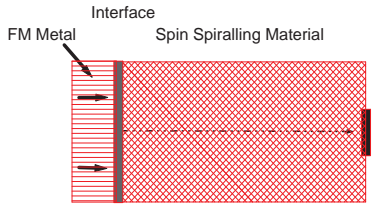}%
\caption{Schematic view of DC spin battery. Arrows in the FM\ metal as a spin
source\ indicate the magnetization is perperdicular to the interface. The spin
spiraling material provides a non-zero spin spirality to drive spin current
along the axis. }%
\label{battery}%
\end{center}
\end{figure}

\textit{A new effective field and spin battery}. To clarify the physical
meaning of the last term, we come to calculate the rate of change of a local
spin. From classical mechanics or from the Heisenberg equation of motion the
rate of change of angular momentum S$_{n}$ is equal to the torque $\mu
_{n}\times\mathbf{h}_{eff}^{n}$ which acts on the spin,%
\begin{equation}
d\mathbf{S}_{n}/dt=-g\mu_{B}\mathbf{S}_{n}\times\mathbf{h}_{eff}^{n},
\end{equation}
where the effective field experienced by the spin is%
\begin{align*}
\mathbf{h}_{eff}^{n}  &  =\frac{t}{g\mu_{B}(2S+1)^{2}}\sum_{\delta}\\
&  \left[  k_{nn+\delta}\mathbf{S}_{n+\delta}+(S+1)\mathbf{M}_{nn+\delta
}+\mathbf{I}_{nn+\delta}^{s}\times\mathbf{S}_{n+\delta}\right]  .
\end{align*}
The first two terms, again are related to particle's orbital motions. The
third term is the torque between the spin current and the local spin.
\textit{A new type of force or effective field is caused due to the
interaction between the spin current and local spin or local magnetization.
}The force may be responsible for magnetization precession by the spin
injection into a magnetic conductor\cite{Weber01} and the anomalous Hall
effect in ferromagnetic materials.\cite{Hirsch99, Niu01} Further discussion
along this line will be published elsewhere. Our observation will lead to
potential applications. For example, by means of the spin spiraling properties
we can design a spin battery to produce a pure spin current. The battery
contains two parts as shown in Fig.\ref{battery}. The main part is a spin
spiraling material. The spins spirals about the axis which is perpendicular to
the interface with a ferromagnetic metal film such that the spin voltage along
the axis is non-zero. The spin chirality in the material will produce the spin
accumulation at two terminals. The magnetization of the film is also
perpendicular to the interface. The film is a source of spins which will flow
out of the battery driven by the spin chirality. The key element in this
battery is the spin spiraling materials. Possible realization of spiral states
have been discussed by several
authors,\cite{Kubo82,Jensen91,Inoue95,Korling96} and the state may also be
stabilized with help of the external magnetic field and exchange coupling
between localized spins.%

In conclusion, we have shown that the strong coupling between the spins of
charge carriers and local spins leads to an effective interaction between spin
current and spin chirality in the system. Non-zero spin chirality may give
rise to a pure spin current and in return a spin current can cause a novel
type of force on the spins. We confirm these findings by studying in detail
the spin transport in the spin spiral state.

The work is supported by grants from the Research Grants Council of Hong Kong,
China (Project HKU7088/01P) and US-DOE.


\begin{thebibliography}{99}
\bibitem {Spiral}As an illustration we consider a ferromagnetic Heisenberg
chain with $n$ spins. The directions of two spins at two terminals can be
controled by strong magnetic fields. Assume the angles between the two spins
are fixed to be $\varphi$. In the classic spin limit the ground state is
obtained by minimizing the classical energy and is shown to be an spin spiral
state: the angle between two nearest neighbour spins is $\varphi/(n-1).$ The
quantum case can be solved by the Bethe ansatz and/or in the spin wave theory.
Other conditions such as longer-range exchange coupling and spin lattice
structures may also give arise a spin spiral state.
\bibitem {Note}The microscopic mechanism should hold in the case of an
antiferromagnetic coupling $J<0$ and large S. In the strong limit the
conduction electron and local spin form a spin $S-1/2$ state. The projection
operator for the state is $\mathbf{P}_{n}^{-}=[-\mathbf{S}_{n}\cdot\tau
+S\tau_{0}]/(2S+1)$ instead of $\mathbf{P}_{n}^{+}.$ The main different point
is that the coupling in the third term in Eq.(\ref{effective}) is
antiferromagnetic instead of ferromagnetic, and the correspondent coefficients
are revised slightly. The ferromagnetism in diluted magnetic semiconductors
originates from the antiferromagnetic coupling between the conduction
electrons and doped magnetic elements.\cite{Dietl00}
\end{thebibliography}
\end{document}